\documentclass[showpacs,showkeys,eqsecnum,twocolumn,prd,aps,twoside]{revtex4-1}

\usepackage{amsfonts,amsmath,amssymb,bm}

\begin{document}

\title{Canonical quantization of a massive Weyl field}

\author{Maxim Dvornikov}
\affiliation{Institute of Physics, University of S\~{a}o Paulo, CP 66318, CEP 05315-970 S\~{a}o Paulo, SP, Brazil}
\email{maxim.dvornikov@usp.br}
\affiliation{Pushkov Institute of Terrestrial Magnetism,
Ionosphere and Radiowave Propagation (IZMIRAN), Troitsk, Moscow
Region 142190, Russia}

\date{\today}

\begin{abstract}
We construct a consistent theory of a quantum massive Weyl field.
We start with the formulation of the classical field theory approach for the description of massive Weyl fields. It is demonstrated that the standard Lagrange formalism cannot be applied for the studies of massive first-quantized Weyl spinors. Nevertheless we show that the classical field theory description of massive Weyl fields can be implemented in frames of the Hamilton formalism or using the extended Lagrange formalism. Then we carry out a
canonical quantization of the system. The independent ways for the quantization of a massive Weyl field are discussed.
We also compare our results with the previous approaches for the treatment of massive Weyl spinors. Finally the new interpretation of the Majorana condition is proposed.
\end{abstract}



\maketitle

\section{Introduction\label{INTR}}

Majorana particles are known to play an important role in the
modern theoretical physics, especially in the studies of
neutrinos. The most natural mechanism for the neutrino mass
generation requires that neutrinos are Majorana
particles~\cite{Kob80,SchVal80}. Although presently there is no
universally recognized experimental results casting light upon the
nature of neutrinos, the attempts are made to investigate
whether neutrinos are Dirac or Majorana particles~\cite{Aug12,And11}. Besides elementary particles physics, Majorana fields can be encountered in the solid states physics. For example, vortices at the interface between an $s$-wave superconductor and the surface of a topological
insulator behave like Majorana particles~\cite{Cha10}.

It is well known that instead of dealing with a four component
spinor $\psi$ satisfying the Majorana condition,
\begin{equation}\label{Majcond}
  \psi^c = \mathrm{i} \gamma^2 \psi^{*} = \varkappa_c \psi,
\end{equation}
the dynamics of a Majorana particle can be re-formulated in terms
of the two component Weyl spinors. Here $\varkappa_c$ is a phase
factor having the unit absolute value. In our analysis we shall
suppose that $\varkappa_c = 1$. We shall use the chiral representation of spinors, in which the Dirac matrices, $\gamma^\mu = (\gamma^0, \bm{\gamma})$, have the form~\cite{ItzZub80},
\begin{equation}\label{Dirmatr}
  \gamma^0 = \left(
               \begin{array}{cc}
                 0 & -I \\
                 -I & 0 \\
               \end{array}
             \right),
  \quad
  \bm{\gamma} = \left(
               \begin{array}{cc}
                 0 & \bm{\sigma} \\
                 -\bm{\sigma} & 0 \\
               \end{array}
             \right),
\end{equation}
where $I$ is the $2 \times 2$ unit matrix and $\bm{\sigma}$ are the
Pauli matrices.

The wave equations for the Weyl
spinors, $\eta$ and $\xi$, have the form,
\begin{equation}\label{weleft}
  \dot{\eta} - c (\bm{\sigma}\nabla)\eta +
  \frac{m c^2}{\hbar} \sigma_2 \eta^{*{}} = 0,
\end{equation}
or
\begin{equation}\label{weright}
  \dot{\xi} + c (\bm{\sigma}\nabla)\xi -
  \frac{m c^2}{\hbar} \sigma_2 \xi^{*{}} = 0,
\end{equation}
where $m$ is the mass of the particle, $c$ is the speed of light, and $\hbar$ is the Planck constant. Note that Eqs.~\eqref{weleft} and~\eqref{weright}
can be formally derived from the Dirac equation,
\begin{equation}\label{Direq}
  \mathrm{i} \hbar \frac{\partial \psi}{\partial t} =
  - \mathrm{i} \hbar c (\bm{\alpha}\nabla)\psi + m c^2 \beta \psi,
\end{equation}
if we suggest that a four component spinor has the form
$\psi_\eta^\mathrm{T} = (\mathrm{i} \sigma_2 \eta^{*{}}, \eta)$ or
$\psi_\xi^\mathrm{T} = (\xi, -\mathrm{i} \sigma_2 \xi^{*{}})$,
which satisfy the Majorana condition~\eqref{Majcond}. The Dirac matrices in Eq.~\eqref{Direq} read
\begin{equation}
  \bm{\alpha} = \gamma^0 \bm{\gamma} =
             \left(
               \begin{array}{cc}
                 \bm{\sigma} & 0 \\
                 0 & -\bm{\sigma} \\
               \end{array}
             \right),
\end{equation}
and $\beta = \gamma^0$, cf. Eq.~\eqref{Dirmatr}. In
the following we will use the natural units in which $\hbar = c =
1$.

Note that we presented the heuristic derivation of
Eqs.~\eqref{weleft} and~\eqref{weright} from Eq.~\eqref{Direq}.
In our analysis we shall just postulate the main
Eqs.~\eqref{weleft} and~\eqref{weright} for the two component Weyl
fields.

It should be noticed that the description of Majorana particles in
terms of the Weyl spinors is more suitable since the electroweak
interaction of elementary particles involves the chiral
projections of four component spinors, $\psi_\mathrm{L,R} = (1 \mp
\gamma^5)/2 \times \psi$, which are equivalent to the Weyl fields
$\eta$ and $\xi$. Here
\begin{equation}
  \gamma^5 = \left(
               \begin{array}{cc}
                 I & 0 \\
                 0 & -I \\
               \end{array}
             \right).
\end{equation}
Despite the equal significance of
Eqs.~\eqref{weleft} and~\eqref{weright}, the former one is
more frequently used for the description of a massive Majorana
neutrino since it was experimentally established that active
neutrinos correspond to left-handed fields~\cite{GolGroSun58}.
That is why we will be mainly interested in Eq.~\eqref{weleft} for
$\eta$. Note that the unitary equivalence of Majorana and Weyl
fields was rigorously proved in Ref.~\cite{FukYan03}.

Before we proceed a remark should be made on the classical field
theory description of a spinor field. The Dirac
equation~\eqref{Direq} contains the Planck constant $\hbar$.
Therefore, besides the case of massless fermions, this wave
equation always corresponds to a quantum particle. However one can
treat the wave function $\psi$ as a $c$-number object and describe
its dynamics in frames of the classical field theory. We will call such a field as classical or first-quantized. One may
speak about a quantized fermion field when $\psi$ is expressed in
terms of the creation and annihilation operators. This terminology
is borrowed from the book by Bogoliubov \& Shirkov~\cite{BogShi80}.

Despite the numerous works devoted to the analysis of
Eq.~\eqref{weleft} are published, still there is a gap in the understanding of
the dynamics of Weyl fields. When one tries to substitute a Majorana spinor $\psi_\eta$ in the Lagrangian for a Dirac field, $\mathcal{L} = \bar{\psi}_\eta (\mathrm{i} \gamma^\mu \partial_\mu - m) \psi_\eta$, one arrives to the following Lagrangian for a Weyl field~\cite{FukYan03}:
\begin{equation}\label{Lagrnaive}
  \mathcal{L} =
  \mathrm{i} \eta^\dag (\sigma^\mu \partial_\mu) \eta -
  \frac{\mathrm{i}}{2} m \eta^\mathrm{T} \sigma_2 \eta +
  \frac{\mathrm{i}}{2} m \eta^\dag \sigma_2 \eta^{*{}},
\end{equation}
where $\sigma^\mu = (I, -\bm{\sigma})$. We, however, notice that the mass term in Eq.~\eqref{Lagrnaive} vanishes if the spinor $\eta$ has commuting $c$-number components, i.e. when $\eta$ is a classical field in our terminology. The solution to this problem was proposed in Ref.~\cite{SchVal81}, where it was suggested that on a classical level a massive Weyl spinor must be described using anticommuting Grassmann variables ($g$-numbers). Thus according to Ref.~\cite{SchVal81} there is no description of massive Weyl particles in terms of the first-quantized fields. However this point of view is in the contradiction with the operator formalism which is commonly used in the quantization of fields~\cite{Wei96}.

The $g$-numbers approach to the treatment of massive Weyl fields was recently criticized in Ref.~\cite{Ahl11}. To construct the $c$-number treatment of Majorana particles the authors of Ref.~\cite{Ahl11} had to introduce special Majorana fields, called \emph{Eigenspinoren des LadungsKonjugationsOperators} (ELKO), which belong to non-standard Wigner classes. The connection of ELKO to the dark matter problem was also studied in Ref.~\cite{Ahl11}.

In the present work we develop a treatment of massive Weyl fields which is based on the standard operator approach for the quantization of fields. Firstly, in Sec.~\ref{CLASSICAL}, we analyze the applicability of the standard Lagrange formalism for the description of massive $c$-number Weyl spinors. Then we propose a classical Hamiltonian for a massive first-quantized Weyl field. The wave equations~\eqref{weleft} and~\eqref{weright} are derived on the basis of this Hamiltonian using the standard variational procedure.
We also show that the extended Lagrange formalism is valid for the description of the evolution of classical massive Weyl fields.
Then, in Sec.~\ref{QUANTUM}, we carry out the canonical
quantization of a massive Weyl field. We find the plane wave solutions of
the wave equations for Weyl fields and calculate their energy
using the Hamiltonian proposed. The requirement of the positive
definiteness of the energy results in the establishment of the
anticommutation expressions for the field amplitudes which turn
out to be operators now. The independent ways of the quantization
of a Weyl field are also considered. Finally, in
Sec.~\ref{CONCLUSION}, we discuss our results.

As we mentioned above the most prominent candidates among fermions to be
described in terms of Majorana fields are neutrinos. It was
experimentally proven that
neutrinos are mixed particles (see Refs.~\cite{An12,Abe12} devoted to the recent achievements in the direct measurement of the mixing angle $\theta_{13}$), whereas the present work is devoted
to the description of a single free Weyl field. Nevertheless the
results of our work can be easily generalized to include several
neutrino generations. Note that the evolution of mixed massive
Dirac and Majorana neutrinos was studied in frames of the
relativistic quantum mechanics (or classical field theory) to
phenomenologically describe neutrino oscillations in vacuum and
various external fields (see the review by Dvornikov~\cite{Dvo11} and references
therein).

\section{Classical field theory\label{CLASSICAL}}

To start with the development of the classical field theory
approach for the description of the massive first-quantized Weyl field $\eta$ we notice that the standard Lagrange formalism does not seem to be a suitable tool for this purpose. The relativistic Euler-Lagrange equation for the field $\eta$ has the form,
\begin{equation}\label{ELe}
  \frac{\partial}{\partial t}
  \frac{\partial \mathcal{L}}{\partial \dot{\eta}} +
  \nabla
  \frac{\partial \mathcal{L}}{\partial \nabla \eta} =
  \frac{\partial \mathcal{L}}{\partial \eta},
\end{equation}
and the analogous equation for $\eta^{*{}}$. We can see that owing to the Lorentz invariance the terms containing the Lagrangian's derivatives with respect to $\dot{\eta}$ and $\nabla \eta$ enter to Eq.~\eqref{ELe} in a symmetric way.

Now we rewrite Eq.~\eqref{weleft} in the equivalent form,
\begin{equation}\label{welefteqiv}
  \sigma_2 \dot{\eta} - \sigma_2 (\bm{\sigma}\nabla)\eta + m \eta^{*{}} = 0.
\end{equation}
Let us discuss two limiting cases: (i) only time dependent spinor, $\nabla \eta = 0$; and (ii) only coordinate dependent spinor, $\dot{\eta} =0$. Now one can see that in the former case the time derivative term in Eq.~\eqref{welefteqiv} is connected to the mass term through the anti-symmetric matrix $\sigma_2$. A classical Lagrangian for such a dynamic equation was discussed in Ref.~\cite{FadJac88}. It should have a kinetic term involving a symplectic two-form. The corresponding evolution equation may be obtained using Eq.~\eqref{ELe} with the following Lagrangian: $\mathcal{L}_t = \tfrac{1}{2} \dot{\eta}^\mathrm{T} \sigma_2 \eta  - \tfrac{1}{2} \dot{\eta}^\dag \sigma_2 \eta^{*{}} - m \eta^\dag \eta$. On the contrary, in the latter case the coordinate derivatives term is connected to the mass term in Eq.~\eqref{welefteqiv} through the symmetric matrices $\sigma_2 \bm{\sigma}$. It means that a classical Lagrangian to obtain such a differential equation should be expressed as $\mathcal{L}_s = - \eta^\dag \sigma_2 (\bm{\sigma}\nabla) \eta + \tfrac{m}{2}(\eta^\dag \eta^{*{}} + \eta^\mathrm{T} \eta)$, cf. Eq.~\eqref{ELe}. The derivatives term of this Lagrangian is a bilinear form with the symmetric matrix.

Now, if one tries to construct a Lagrangian for the general case when neither $\nabla \eta$ nor $\dot{\eta}$ are equal to zero, one notices that it is impossible to reconcile the structure of Eq.~\eqref{welefteqiv} with the relativistic invariant Eq.~\eqref{ELe} in case of a $c$-number spinor $\eta$. Indeed varying the Lagrangian $\mathcal{L}_t$, according to Eq.~\eqref{ELe}, with respect to $\eta$ we get a differential equation for $\eta$. However, if we vary the Lagrangian $\mathcal{L}_s$ with respect to $\eta$ we obtain a differential equation for $\eta^{*{}}$. Thus a general classical Lagrangian, bilinear in fields $\eta$ and $\eta^{*{}}$, is unlikely to be constructed.

The above heuristic discussion shows that the standard Lagrange formalism cannot be applied for the studies of massive classical (first-quantized) Weyl fields. However, the Lagrange formalism is not a unique way to obtain a wave equation with help of a variational procedure. We can instead discuss a Hamilton approach for the studies of massive classical Weyl fields. Let us consider the following Hamiltonian:
\begin{align}\label{Hamclass}
  H[\eta,\eta^{*{}},\pi,\pi^{*{}}] = & \int \mathrm{d}^3\mathbf{r}
  \big\{
    \pi^\mathrm{T} (\bm{\sigma}\nabla) \eta -
    (\eta^{*{}})^\mathrm{T} (\bm{\sigma}\nabla) \pi^{*{}}
    \notag
    \\
    & +
    m
    \left[
      (\eta^{*{}})^\mathrm{T} \sigma_2 \pi +
      (\pi^{*{}})^\mathrm{T} \sigma_2 \eta
    \right]
  \big\},
\end{align}
which is a functional of independent canonical variables $(\eta,\pi)$ and $(\eta^{*{}},\pi^{*{}})$. Taking into account that the particle's mass, $m$, must be a real parameter, we get that $H$ is also real as it should be for a classical Hamiltonian.
%
%

Using the classical field theory version of the canonical Hamilton
equations,
\begin{align}\label{etaclass}
  \dot{\eta}  = & \frac{\delta H}{\delta \pi} =
  (\bm{\sigma}\nabla)\eta - m \sigma_2 \eta^{*{}},
  \notag
  \\
  \dot{\eta}^{*{}}  = & \frac{\delta H}{\delta \pi^{*{}}} =
  (\bm{\sigma}^{*{}}\nabla)\eta^{*{}} + m \sigma_2 \eta,
\end{align}
we obtain Eq.~\eqref{weleft} for a massive Weyl field. With help
of the second pair of the canonical equations,
\begin{align}\label{piclass}
  \dot{\pi}  = & - \frac{\delta H}{\delta \eta} =
  (\bm{\sigma}^{*{}}\nabla)\pi + m \sigma_2 \pi^{*{}},
  \notag
  \\
  \dot{\pi}^{*{}} = & - \frac{\delta H}{\delta \eta^{*{}}} =
  (\bm{\sigma}\nabla)\pi^{*{}} - m \sigma_2 \pi,
\end{align}
one gets the equations for the canonical momenta. If we introduce
the new variables $\xi = \mathrm{i} \sigma_2 \pi$ and $\xi^{*{}} =
\mathrm{i} \sigma_2 \pi^{*{}}$, Eq.~\eqref{piclass} becomes
equivalent to Eq.~\eqref{weright} for $\xi$.

Note that Eq.~\eqref{etaclass} for ``coordinates" does not
contain momenta and vice versa. Thus two groups of variables
$(\eta, \eta^{*{}})$ and $(\pi, \pi^{*{}})$ evolve in time
independently. It means that one cannot find the relation between
the canonical momenta and the ``velocities", $\dot{\eta}$ and
$\dot{\eta}^{*{}}$, to construct a Lagrangian~\cite{Gan75}. This fact will be discussed in Sec.~\ref{CONCLUSION} in details.


Despite that no conventional Lagrange formalism can be applied for the description of our system, we can construct an extended Lagrangian, $\tilde{\mathcal{L}}$, which also includes the momenta, $\pi$ and  $\pi^{*{}}$, as well as their time derivatives, $\dot{\pi}$ and $\dot{\pi}^{*{}}$, as independent variables. Let us choose the extended Lagrangian as~\cite{FadJac88}
\begin{align}\label{extLagr}
  \tilde{\mathcal{L}} = &
  \pi^\mathrm{T} \dot{\eta} + (\pi^{*{}})^\mathrm{T} \dot{\eta}^{*{}} -
  \frac{1}{2}
  \left[
    \pi^\mathrm{T} (\bm{\sigma}\nabla) \eta -
    (\eta^{*{}})^\mathrm{T} (\bm{\sigma}\nabla) \pi^{*{}}
  \right]
  \notag
  \\
  & -
  m
  \left[
    (\eta^{*{}})^\mathrm{T} \sigma_2 \pi +
    (\pi^{*{}})^\mathrm{T} \sigma_2 \eta
  \right].
\end{align}
Varying this Lagrangian with respect to $\eta$ or $\eta^{*{}}$ and using Eq.~\eqref{ELe}, we get the wave equations for $\pi$ or $\pi^{*{}}$, cf. Eq.~\eqref{piclass}. Making the same variational procedure with the independent variables $\pi$ or $\pi^{*{}}$ one can reproduce Eq.~\eqref{etaclass}. Again we can see that two groups of variables, $(\eta,\eta^{*{}})$ and $(\pi,\pi^{*{}})$, evolve independently.

It should be noted that the extended Lagrangian $\tilde{\mathcal{L}}$ is not equal to $\tilde{\mathcal{L}}' = \pi^\mathrm{T} \dot{\eta} + (\pi^{*{}})^\mathrm{T} \dot{\eta}^{*{}} - \mathcal{H}$, where $\mathcal{H}$ is the Hamiltonian density, as one can expect from the standard approach~\cite{GitTyu90}. In fact there is an additional factor $1/2$ in the spatial derivatives term in Eq.~\eqref{extLagr}. One can say that the evolution of the system, based on Eq.~\eqref{extLagr}, is an extended Lagrange dynamics in the analogy to the extended Hamilton formalism~\cite{GitTyu90}.


\section{Quantization\label{QUANTUM}}

Using the results of Ref.~\cite{FukYan03} we find the plane wave
solutions of Eqs.~\eqref{weleft} and~\eqref{weright} in the
following form:
\begin{align}\label{etaxisol}
  \eta(x) = & \frac{1}{2}
  \int \frac{\mathrm{d}^3\mathbf{p}}{(2\pi)^{3/2}}
  \sqrt{1 + \frac{E}{|\mathbf{p}|}}
  \notag
  \\
  & \times
  \bigg[
    \left(
      a_{-{}} w_{-{}} -
      \frac{m}{E+|\mathbf{p}|} a_{+{}} w_{+{}}
    \right) e^{-\mathrm{i}px}
    \notag
    \\
    & +
    \left(
      a_{+{}}^{*{}} w_{-{}} +
      \frac{m}{E+|\mathbf{p}|} a_{-{}}^{*{}} w_{+{}}
    \right) e^{\mathrm{i}px}
  \bigg],
  \notag
  \displaybreak[2]
  \\
  \xi(x) = & \frac{\mathrm{i}}{2}
  \int \frac{\mathrm{d}^3\mathbf{p}}{(2\pi)^{3/2}}
  \sqrt{1 + \frac{E}{|\mathbf{p}|}}
  \notag
  \\
  & \times
  \bigg[
    \left(
      b_{+{}} w_{+{}} +
      \frac{m}{E+|\mathbf{p}|} b_{-{}} w_{-{}}
    \right) e^{-\mathrm{i}px}
    \notag
    \\
    & +
    \left(
      b_{-{}}^{*{}} w_{+{}} -
      \frac{m}{E+|\mathbf{p}|} b_{+{}}^{*{}} w_{-{}}
    \right) e^{\mathrm{i}px}
  \bigg],
\end{align}
where $p^\mu = (E,\mathbf{p})$, $E = \sqrt{\mathbf{p}^2+m^2}$ is
the energy of a particle, and $w_\sigma$, $\sigma = \pm{}$, are
the helicity amplitudes. Here we list some of their useful
properties,
\begin{gather}\label{wprop}
  (\bm{\sigma}\mathbf{p}) w_\sigma = \sigma |\mathbf{p}| w_\sigma,
  \quad
  \mathrm{i} \sigma_2 w_\sigma^{*{}} = -\sigma w_{-\sigma},
  \notag
  \\
  w_\sigma(-\mathbf{p}) = \mathrm{i} w_{-\sigma}(\mathbf{p}).
\end{gather}
We can choose $w_\sigma$ in the explicit form as~\cite{BerLifPit82}
\begin{align}
  w_{+{}} = & \left(
              \begin{array}{c}
                e^{-\mathrm{i}\phi/2} \cos \theta/2 \\
                e^{\mathrm{i}\phi/2} \sin \theta/2 \\
              \end{array}
            \right),
  \notag
  \\
  w_{-{}} = & \left(
              \begin{array}{c}
                -e^{-\mathrm{i}\phi/2} \sin \theta/2 \\
                e^{\mathrm{i}\phi/2} \cos \theta/2 \\
              \end{array}
            \right),
\end{align}
where $\phi$ and $\theta$ are the angles giving the direction of the momentum
of a particle, $\mathbf{p} = |\mathbf{p}|(\sin \theta \cos \phi, \sin \theta \sin \phi, \cos \theta)$.

In the classical field theory~\cite{Dvo11}, the
expansion coefficients $a_{\pm{}}(\mathbf{p})$ and
$b_{\pm{}}(\mathbf{p})$ were supposed to be $c$-number functions.
However now we assume that these objects are commuting or
anticommuting operators. The type of statistics will be chosen to
provide the positive definiteness of the energy. It should be also
noted that we take the different kinds of operators in the
decomposition of $\eta$ and $\xi$ since, as we
mentioned above, these fields evolve independently.

On the basis of Eqs.~\eqref{Hamclass}, \eqref{etaxisol},
and~\eqref{wprop}, after a bit lengthy but straightforward
calculations we get the Hamiltonian expressed in terms of the
operators $a_{\pm{}}(\mathbf{p})$ and $b_{\pm{}}(\mathbf{p})$ and
their conjugate,
\begin{widetext}
\begin{align}\label{Hamquant}
  H = & \frac{1}{4} \int \mathrm{d}^3\mathbf{p}
  E
  \left(
    1 + \frac{E}{|\mathbf{p}|}
  \right)
  \bigg\{
  \Big\{
    a_{-{}}^{*{}}(\mathbf{p}) b_{-{}}(\mathbf{p}) +
    b_{-{}}^{*{}}(\mathbf{p}) a_{-{}}(\mathbf{p}) -
    a_{+{}}(\mathbf{p}) b_{+{}}^{*{}}(\mathbf{p}) -
    b_{+{}}(\mathbf{p}) a_{+{}}^{*{}}(\mathbf{p})
    \notag
    \\
    & +
    \left(
      \frac{m}{E+|\mathbf{p}|}
    \right)^2
    \left[
      a_{-{}}(\mathbf{p}) b_{-{}}^{*{}}(\mathbf{p}) +
      b_{-{}}(\mathbf{p}) a_{-{}}^{*{}}(\mathbf{p}) -
      a_{+{}}^{*{}}(\mathbf{p}) b_{+{}}(\mathbf{p}) -
      b_{+{}}^{*{}}(\mathbf{p}) a_{+{}}(\mathbf{p})
    \right]
    \Big\}
    \notag
    \displaybreak[2]
    \\
    & +
    \mathrm{i}
    \frac{m}{|\mathbf{p}|}
    \Big\{
      e^{-2\mathrm{i}Et}
      \left[
        a_{-{}}(\mathbf{p}) b_{-{}}(-\mathbf{p}) +
        b_{-{}}(-\mathbf{p}) a_{-{}}(\mathbf{p}) +
        b_{+{}}(-\mathbf{p}) a_{+{}}(\mathbf{p}) +
        a_{+{}}(\mathbf{p}) b_{+{}}(-\mathbf{p})
      \right]
      \notag
      \\
      & +
      e^{2\mathrm{i}Et}
      \left[
        a_{-{}}^{*{}}(\mathbf{p}) b_{-{}}^{*{}}(-\mathbf{p}) +
        b_{-{}}^{*{}}(-\mathbf{p}) a_{-{}}^{*{}}(\mathbf{p}) +
        b_{+{}}^{*{}}(-\mathbf{p}) a_{+{}}^{*{}}(\mathbf{p}) +
        a_{+{}}^{*{}}(\mathbf{p}) b_{+{}}^{*{}}(-\mathbf{p})
      \right]
    \Big\}
  \bigg\}.
\end{align}
\end{widetext}
Now we establish the following relation between the independent
operators $a_{\pm{}}(\mathbf{p})$ and $b_{\pm{}}(\mathbf{p})$:
\begin{equation}\label{Majcondoper}
  a_{\pm{}}(\mathbf{p}) = b_{\pm{}}(\mathbf{p}),
\end{equation}
and the analogous expression for the conjugate operators. We will
choose the operators $a_{\pm{}}(\mathbf{p})$ as the basic ones and
assume that they obey the anticommutation relations,
\begin{equation}\label{anticomrel}
  \{
    a_\sigma(\mathbf{k});a_{\sigma'}^{*{}}(\mathbf{p})
  \}_{+{}} =
  \delta_{\sigma\sigma'} \delta^3(\mathbf{k} - \mathbf{p}),
\end{equation}
with all the other anticommutators being equal to zero. In this
case the time dependent terms in Eq.~\eqref{Hamquant} are washed
out. Using Eqs.~\eqref{Majcondoper} and~\eqref{anticomrel} we can
recast Eq.~\eqref{Hamquant} into the form
\begin{equation}\label{totenquant}
  H = \int \mathrm{d}^3\mathbf{p}
  \thinspace E
  (a_{-{}}^{*{}} a_{-{}} +
  a_{+{}}^{*{}} a_{+{}}) + \text{divergent terms},
\end{equation}
which shows that the total energy of a Weyl field is a sum of the
energies of elementary oscillators corresponding to the negative
and the positive helicity states.

In the canonical formalism the total momentum of our system can be
calculated using the expression,
\begin{equation}
  \mathbf{P} = \int \mathrm{d}^3\mathbf{r}
  \left[
    (\eta^{*{}})^\mathrm{T} \nabla \pi^{*{}} -
    \pi^\mathrm{T} \nabla \eta
  \right],
\end{equation}
which is obtained by the spatial integration of the $T^{i0}$
component of the energy-momentum tensor $T^{\mu\nu}$. Omitting the
detailed calculations and with help of Eqs.~\eqref{etaxisol},
\eqref{wprop}, \eqref{Majcondoper}, and~\eqref{anticomrel} we get
the following formula for the quantized momentum of the Weyl
field:
\begin{equation}\label{totmomquant}
  \mathbf{P} = \int \mathrm{d}^3\mathbf{p}
  \thinspace
  \mathbf{p} (a_{-{}}^{*{}} a_{-{}} +
  a_{+{}}^{*{}} a_{+{}}) + \text{divergent terms},
\end{equation}
which has the analogous structure as Eq.~\eqref{totenquant}.

There is, however, another way to quantize a Weyl field. Instead
of Eq.~\eqref{Majcondoper} we may choose the following relation
between the operators:
\begin{equation}\label{Majcondoperalt}
  a_{\pm{}}(\mathbf{p}) = b_{\mp{}}(\mathbf{p}),
\end{equation}
with the condition~\eqref{anticomrel} still being held true for
the operators $a_{\pm{}}$. In this case the time dependent terms
in Eq.~\eqref{Hamquant} are also equal to zero. To diagonalize the
remaining time independent expression in Eq.~\eqref{Hamquant} we
introduce the new operators $c_{\pm{}}$ by means of the Bogoliubov
transformation,
\begin{equation}\label{physdegfreedom}
  a_{-{}}  = \frac{1}{\sqrt{2}}(c_{-{}} - c_{+{}}^{*{}}),
  \quad
  a_{+{}}  = \frac{1}{\sqrt{2}}(c_{-{}} + c_{+{}}^{*{}}).
\end{equation}
One can check by a direct calculation that the new operators also
satisfy the canonical anticommutation properties: $\{ c_\sigma(\mathbf{k}); c_{\sigma'}^{*{}}(\mathbf{p}) \}_{+{}} =
\delta_{\sigma\sigma'} \delta^3(\mathbf{k} - \mathbf{p})$, etc. Finally, the
secondly quantized Hamiltonian is expressed as
\begin{equation}\label{totenquantalt}
  H = \int \mathrm{d}^3\mathbf{p}
  \thinspace E
  (c_{-{}}^{*{}} c_{-{}} +
  c_{+{}}^{*{}} c_{+{}}) + \text{divergent terms}.
\end{equation}
One can also show that, after the quantization in terms of the
operators $c_{\pm{}}$, the total momentum takes the form,
\begin{equation}\label{totmomquantalt}
  \mathbf{P} = \int \mathrm{d}^3\mathbf{p}
  \thinspace
  \mathbf{p} (c_{-{}}^{*{}} c_{-{}} +
  c_{+{}}^{*{}} c_{+{}}) + \text{divergent terms}.
\end{equation}
We can see that Eqs.~\eqref{totenquantalt}
and~\eqref{totmomquantalt} have the same structure as
Eqs.~\eqref{totenquant} and~\eqref{totmomquant} respectively.

\section{Summary and discussion\label{CONCLUSION}}

In summary we mention that in the present work we have carried out
a consistent canonical quantization of a massive Weyl field. Two
major results have been obtained.

Firstly, in Sec.~\ref{CLASSICAL},
we have constructed a classical field theory approach for the
description of the massive Weyl field dynamics. The classical field theory
was applied in the form of the canonical Hamilton formalism since it has been demonstrated that a standard Lagrangian, bilinear in the independent classical fields $\eta$ and $\eta^{*{}}$ and their time derivatives, is unlikely to exist. On the basis of the
proposed classical Hamiltonian~\eqref{Hamclass} and using a
standard variational procedure we have obtained the main
Eqs.~\eqref{weleft} and~\eqref{weright} for massive Weyl spinors. We also constructed the extended Lagrangian~\eqref{extLagr}, which includes the momenta and their derivatives as independent variables (see Refs.~\cite{FadJac88,GitTyu90}). In frames of the extended Lagrange formalism we could reproduce the wave equations~\eqref{weleft} and~\eqref{weright}.

Note that a consistent field theory treatment of massive Weyl fields faced certain difficulties since the first attempt to quantize them has been made in Ref.~\cite{Cas57}. As we have mentioned in Sec.~\ref{INTR}, the direct application of the Lagrange formalism for Dirac fields to the description of a Majorana spinor $\psi_\eta$ leads to the Lagrangian~\eqref{Lagrnaive} for a two component spinor $\eta$. The Lagrangian~\eqref{Lagrnaive} has a mass term vanishing when $\eta$ is a $c$-number classical field. To resolve this issue in Ref.~\cite{Cas57} it was suggested that $\eta$ should be already expressed via anticommuting operators. However, from the logical point of view, such a treatment is just a re-expression of already quantized objects in terms of new variables rather than a generic quantization.

Another solution, how to save the mass term in Eq.~\eqref{Lagrnaive}, has been suggested in Ref.~\cite{SchVal81}. In that work it was proposed that a classical massive Weyl field must be expressed via anticommuting $g$-numbers. Nevertheless, if one claims that the $g$-numbers description is a unique representation of a classical field, it is in contradiction with the standard operator formulation of the quantization procedure, in which a classical $c$-number field is required. The third solution of the puzzle of the mass term in Eq.~\eqref{Lagrnaive} was recently suggested in Ref.~\cite{Ahl11}. Criticizing the $g$-numbers approach for the description of a massive Weyl field, the authors of Ref.~\cite{Ahl11} introduced a special $1/2$-spin field, ELKO, which possesses Majorana properties.

Presently the $g$-numbers description of classical fermions in frames of the Lagrange formalism is commonly used. It is related to the path-integral formulation of the quantum field theory where a classical action is required. For example, the $g$-numbers treatment of pseudoclassical massless Weyl fields was elaborated in Ref.~\cite{GitGonTyu94}. Nevertheless the attempts to develop a classical field theory description of fermion fields, based on $c$-number variables, have been made previously. Besides the aforementioned work by Ahluwalia \textit{et al.}~\cite{Ahl11}, one can recall the studies of Barut \& Zanghi~\cite{BarZan84} devoted to the development of the classical field theory of an electron in an external electromagnetic field.

Unlike Schechter \& Valle~\cite{SchVal81}, who suggested that the $g$-numbers approach is the unique treatment of a classical massive Weyl field, in the present work we have demonstrated that the dynamics of such a field
can be perfectly described using $c$-number fields $\eta$ and $\eta^{*{}}$. However for this purpose we had to use the Hamilton or the extended Lagrange formalisms. In Sec.~\ref{CLASSICAL} we have demonstrated that these approaches are valid for for the description of a classical massive Weyl field. The Lagrange formalism in its standard formulation cannot be applied to study the dynamics of this system.

It is known that the standard
Lagrange and the Hamilton formalisms for the treatment of
classical particles are almost equivalent: for the existing
Lagrangian one can always construct a Hamiltonian, but the
opposite statement is not true. The example of the real physical
process, the rays of light propagation in a medium, which can be
described only within the canonical formalism since the Lagrangian
for such a system is trivial, $L = 0$, is given in Ref.~\cite{LanLif94}.
Note that Hamiltonians analogous to
Eq.~\eqref{Hamclass}, resulting in the first order evolution
equations, were discussed in Ref.~\cite{ZakKuz97} for the studies
of nonlinear waves in frames of the Hamilton formalism.

Recently Dvornikov \& Maalampi~\cite{DvoMaa09} used classical solutions of Eq.~\eqref{weleft} to phenomenologically describe
Majorana neutrino oscillations in vacuum and in external fields. Now the classical field
theory approach for the treatment of Weyl spinors is fully
substantiated. Although we do not doubt that our world is quantum,
numerous processes may be also described within the classical
physics. Many interesting examples of such situations are presented in Ref.~\cite{Gui03}.

%

The second important result obtained in the present work is the
new interpretation of the Majorana condition. Typically Eq.~\eqref{Majcond} is interpreted as the equality of ``particle" and ``antiparticle" degrees of freedom. We should notice that, from the point of view of the quantum field theory, ``particles" and ``antiparticles" may be well defined only after the quantization of a system. Thus it would be reasonable to apply the Majorana condition~\eqref{Majcond} after the fields quantization rather than before it as it was made in Ref.~\cite{FukYan03}.

In Sec.~\ref{INTR} we have mentioned that the two component spinor $\eta$ corresponds to the left-handed chiral projection of the four component spinor. Thus, if we deal, e.g., with neutrino fields, the spinor $\eta$ can be regarded as the``particle" degree of freedom. Using the same argumentation we can relate the spinor $\pi$, which is proportional to the right-handed chiral projection, to the ``antiparticle" degree of freedom. As we have shown in Sec.~\ref{CLASSICAL}, the fields $\eta$ and $\pi$ evolve independently on the classical level. While quantizing the system in Sec.~\ref{QUANTUM}, we had to establish the connection between the degrees of freedom corresponding to $\eta$ and $\pi$. It means that Eqs.~\eqref{Majcondoper} and~\eqref{Majcondoperalt} may be regarded as quantum Majorana conditions. Note that these conditions are applied after the quantization of the system, as it should be.

As in Ref.~\cite{Ahl11}, in the present work we have also used the conventional operator approach to quantize massive Weyl fields. However, contrary to Ref.~\cite{Ahl11}, in our method it was not necessary to introduce any exotic fields, like ELKO. We have obtained the Lorentz invariant wave equations~\eqref{etaclass} and~\eqref{piclass} in frames of our approach and constructed classical and quantum dynamics of these fields. It means that our treatment of massive Weyl fields implies Lorentz invariance unlike the approach involving ELKO spinors~\cite{Ahl11}.

A remark on the weight coefficient in Eq.~\eqref{etaxisol} should
be made. One notices that the function $\rho(\mathbf{p}) = \sqrt{1
+ E/|\mathbf{p}|}$ becomes singular at $|\mathbf{p}| \to 0$. We
should however notice that $\rho$ is a square-integrable function.
Indeed, for $m \neq 0$ we get
\begin{equation}
  4\pi \int_0^{\ldots{}}
  \mathbf{p}^2 \mathrm{d}|\mathbf{p}|\ \rho^2(\mathbf{p}) < \infty.
\end{equation}
Thus the Fourier transform in Eq.~\eqref{etaxisol} is well
defined.
We also mention that in solving of linear problems, e.g., related
to oscillations of Majorana neutrinos, cf.~Ref.~\cite{Dvo11}, one
may choose a non-singular weight coefficient like in the book by Fukugita \& Yanagida~\cite{FukYan03}.

\begin{acknowledgments}
I am very thankful to D.M.~Gitman, J.~Lukierski, and J.~Maalampi for
helpful discussions, to S.~Forte for bringing Ref.~\cite{FadJac88} to my attention, as well as to FAPESP (Brazil) for a grant.
\end{acknowledgments}


\begin{thebibliography}{40}

\bibitem{Kob80}
  Kobzarev,~I.Yu., Martem'yanov,~B.V.,  Okun',~L.B., Shchepkin,~M.G.:
  The phenomenology of neutrino oscillations.
   Sov. J. Nucl. Phys. \textbf{32}, 823--828 (1980)

\bibitem{SchVal80}
  Schechter,~J., Valle~J.W.F.:
  Neutrino masses in $\mathrm{SU}(2) \otimes \mathrm{U}(1)$ theories.
  Phys. Rev. D \textbf{22}, 2227--2235 (1980)

\bibitem{Aug12}
  Auger, M., et al. (EXO Collaboration):
  Search for neutrinoless double-beta decay in $^{136}\mathrm{Xe}$ with EXO-200.
  Phys. Rev. Lett. \textbf{109}, 032505 (2012).
  arXiv:1205.5608~[hep-ex]


\bibitem{And11}
  Andreotti,~E., et al. (CUORICINO Collaboration):
  $^{130}\mathrm{Te}$ neutrinoless double-beta decay with CUORICINO.
  Astropart. Phys. \textbf{34}, 822--831 (2011).
  arXiv:1012.3266~[nucl-ex]



\bibitem{Cha10}
  Chamon,~C., Jackiw,~R., Nishida,~Y., Pi,~S.-Y., Santos,~L.:
  Quantizing Majorana fermions in a superconductor.
  Phys. Rev. B \textbf{81}, 224515 (2010).
  arXiv:1001.2760~[cond-mat.str-el].

\bibitem{ItzZub80}
  Itzykson,~C., Zuber~J.-B.:
  Quantum field theory,
  p.~694.
  McGraw-Hill, New York (1980)

\bibitem{GolGroSun58}
  Goldhaber,~M. Grodzins,~L., Sunyar,~A.W.:
  Helicity of neutrinos,
  Phys. Rev. \textbf{109}, 1015--1017 (1958)

\bibitem{FukYan03}
  Fukugita,~M., Yanagida,~T.:
  Physics of neutrinos and applications to astrophysics,
  pp.~289--319.
  Springer, Berlin (2003)

\bibitem{BogShi80}
  Bogoliubov,~N.N., Shirkov,~D.V.:
  Introduction to the theory of quantized fields,
  3rd ed. pp.~10--89. 
  Wiley, New York (1980)

\bibitem{SchVal81}
  Schechter,~J., Valle,~J.W.F.:
  Majorana neutrinos and magnetic fields.
  Phys. Rev. D \textbf{24}, 1883--1889 (1981)

\bibitem{Wei96}
  Weinberg,~S.:
  The quantum theory of fields: foundations,
  2nd ed. pp.~292--338.
  Cambridge University Press, Cambridge (1996)

\bibitem{Ahl11}
  Ahluwalia,~D.V., Lee,~C.-Y., Schritt, D.:
  Self-interacting Elko dark matter with an axis of locality.
  Phys. Rev. D \textbf{83}, 065017 (2011).
  arXiv:0911.2947~[hep-ph]

\bibitem{An12}
  An,~F.P., et al. (Daya Bay Collaboration):
  Observation of electron-antineutrino disappearance at Daya Bay.
  Phys. Rev. Lett. \textbf{108}, 171803 (2012).
  arXiv:1203.1669~[hep-ex].

\bibitem{Abe12}
  Abe,~Y., et al. (Double Chooz Collaboration):
  Indication of reactor $\bar{\nu}_e$ disappearance in the Double Chooz experiment.
  Phys. Rev. Lett. \textbf{108}, 131801 (2012).
  arXiv:1112.6353~[hep-ex].

\bibitem{Dvo11}
  Dvornikov,~M.:
  Field theory description of neutrino oscillations.
  In:
  Greene,~J.P., (ed.)
  Neutrinos: properties, sources and detection,
  pp.~23--90.
  NOVA Science Publishers, New York (2011).
  arXiv:1011.4300~[hep-ph]

\bibitem{FadJac88}
  Faddeev,~L., Jackiw,~R.:
  Hamiltonian reduction of unconstrained and constrained systems.
  Phys. Rev. Lett. \textbf{60}, 1692--1694 (1988).


\bibitem{Gan75}
  Gantmacher,~F.:
  Lectures in analytical mechanics,
  pp.~71--80.
  Mir Publishers, Moscow (1975)


\bibitem{GitTyu90}
  Gitman,~D.M., Tyutin,~I.V.:
  Quantization of fields with constraints,
  pp.~13--21.
  Springer, Berlin (1990)

\bibitem{BerLifPit82}
  Berestetski\u{\i},~V.B., Lifshitz,~E.M., Pitaevski\u{\i},~L.P.:
  Quantum electrodynamics,
  2nd ed. p.~86.
  Pergamon, Oxford (1980)

\bibitem{Cas57}
  Case,~K.M.:
  Reformulation of the Majorana theory of the neutrino.
  Phys. Rev. \textbf{107}, 307--316 (1957)


\bibitem{GitGonTyu94}
  Gitman,~D.M., Gon\c{c}alves,~A.E., Tyutin,~I.V.:
  New pseudoclassical model for Weyl particles.
  Phys. Rev. D \textbf{50}, 5439--5442 (1994)

\bibitem{BarZan84}
  Barut,~A.O., Zanghi,~N.:
  Classical model of the Dirac electron.
  Phys. Rev. Lett. \textbf{52}, 2009--2012 (1984)

\bibitem{LanLif94}
  Landau,~L.D., Lifshitz,~E.M.:
  The classical theory of fields,
  4th ed. pp.~140--143.
  Butterworth-Heinemann, Amsterdam (1994)

\bibitem{ZakKuz97}
  Zakharov, V.E., Kuznetsov, E.A.:
  Hamiltonian formalism for nonlinear waves.
  Phys. Usp. \textbf{40}, 1087--1116 (1997)

\bibitem{DvoMaa09}
  Dvornikov,~M., Maalampi,~J.:
  Oscillations of Dirac and Majorana neutrinos in matter and a magnetic
  field.
  Phys. Rev. D \textbf{79}, 113015 (2009).
  arXiv:0809.0963~[hep-ph]

\bibitem{Gui03}
  Joos,~E., Zeh,~H.D., Kiefer,~C., Giulini,~D.J.W., Kupsch,~J., Stamatescu,~I.-O.:
  Decoherence and the appearance of a classical world in quantum
  theory,
  2nd ed.
  Springer, Berlin (2003)

\end{thebibliography}
\end{document}